# Transmutation elements Re/Ta effect on vacancy formation and dissociation behavior in W

Shulong Wen[1], Kaige Hu[2], Min Pan[1*], Zheng Huang[3], Zelin Cao[3], Yini Lv[3], Yong Zhao[1]

[1] *Key Laboratory of Advanced Technology of Materials (Ministry of Education), Superconductivity and New Energy R&D Center, Southwest JiaoTong University, Chengdu, Sichuan, 610031, China.*

[2] *School of Physics and Optoelectronic Engineering, Guangdong University of Technology, Guangzhou, Guangdong, 510006, China*

[3] *School of Physical Science Technology, Southwest Jiaotong University, Chengdu, Sichuan, 610031, China.*

Transmutation elements are the essential products in plasma facing materials tungsten, and have further effects on point defects evolution resulted by radiation. Here, transmutation elements Re and Ta atoms have been selected to assess the effects on property of vacancy and vacancy cluster in W material via first-principles calculations. The formation energy indicates mono-vacancy is more likely to form in W-Re system than pure W and W-Ta system. Both Re and Ta have reduced the diffusion barrier energy in the mono-vacancy migration. The calculation presents that vacancy cluster prefers to grow up by combining another vacancy cluster relative to a single mono-vacancy. Re is favorable to the nucleation and growth of vacancies clusters, while Ta has a suppressive effect on the aggregation of small vacancy cluster. The emphasis analysis is obtained according the volumetric dependent strain. Vacancy dissociation calculations show that the dissociation of vacancy clusters is easier to begin with a single vacancy dissociation process.

## 1. Introduction

Tungsten (W) has been considered as one of the most promising candidates for plasma-facing materials (PFMs) in future fusion reactors, due to its particular characteristics including high melting point, high thermal conductivity and low sputtering erosion [1-3]. High energy neutrons (~14 MeV) and high fluxes of ions would be generated in a fusion reaction. The irradiation of such particles would produce various defects in the PFMs, including self-interstitial atoms (SIAs), mono-vacancy, vacancy cluster, transmutation, etc. The diffusion of defects at operating temperature would change the microstructure, and further macroscopically degrade the mechanical properties and service performance of PFMs [4,5].

Previous researches suggest that the primary transmutation product of W under neutron irradiation is rhenium (Re) [6], and the concentration of Re would reach up to 11.9 at. % after five years of exposure to fusion neutrons [7]. However, theoretical studies predicated the concentrations of Re and Ta atoms reach up to 3.8 at.% and 0.81 at.%, respectively [8]. Besides these, transmutation elements (TE) can possibly play an important role to low temperature brittleness, recrystallization behavior, and radiation stability to W lattice. For example, the brittle-to-ductile transition temperature of W would reduce and the ductility and hardness would increase with the Re and Os adoption [9]. Re with W would also reduce the radiation swelling of W [10]. 2MeV W$^+$ ion irradiation induced the formation the Re-rich clusters in the binary W-2 at. % Re alloy and which will cause likely irradiation hardening [11]. The mechanism of radiation-induced has been explored by first-principles method [12].

However, Ta does not shows clustering behavior after irradiation and its presence reduced the Re cluster number density and volume fraction [13]. In addition, experiments showed that solid W-Ta alloys exhibit deviations from the ideal solid solution behavior over the temperature between 1050 and 1200 K [14].

The behavior of point defects can be affected by the transmutation elements which acts as alloying in W. The mono-vacancy formation energy is highly sensitive to the local environment, and it should be between 3eV to 5eV varying from the alloying element composition and the mono-vacancy sites [15-16]. Effects of alloying transmutation impurities on the stability and mobility of helium (He) indicate that the diffusion barrier energy of He around the possible alloying elements is Ta>Re, and Ta might have been chosen as a relatively suitable alloying element compared with other possible ones [17].

The above results are consistent with earlier first-principles calculations in body-centered cubic (bcc) W [18-19]. Isochronal anneals suggest that the principal recombination regions were observed at about 0.15 $T_m$, 0.22 $T_m$ and 0.31 $T_m$ ($T_m$ is the tungsten melting temperature) [20], which shows a phased stage for the point defect recovery. In addition, the recovery curve for annealed W indicates that the stage III in the quenching process is generally attributed to the vacancies' migration with activation energy of 1.7 eV at 700 K. The calculation model of the mono-vacancy diffusion in pure W represents that multiple nearest neighbor jumping styles play an important role in the diffusion coefficient, especially at the temperature over 2/3 of $T_m$ [21]. Nevertheless, the vacancy diffusion in W-alloys should be investigated further. A recent research reveals that the void formation and evolution in Ta-W alloys are closely relative to the solute content [22]. Compared to pure Ta, W atom delays the occurring of vacancy, serving as the solute effect. Despite their relatively high resistance to void swelling of bcc materials, the defect agglomerates such as voids evolution should be systematically assessed and predicted reliably in W-based alloys for their use in future nuclear reactor systems.

In this paper, first-principles methods were used to investigate the vacancy behaviors and their correlations with the transmutation elements in W. To provide a theoretical mechanism on the influence of transmutation elements Re/Ta on the defect evolution, the model for the mono-vacancy migration and clustering properties were examined correspondingly. We found that the vacancy diffusion mainly depend on their migration paths in W-TE system. The formation and diffusion of a vacancy is expected to be favored when there is a Re/Ta in its neighborhood. Nevertheless, contrast to the Re atom, Ta acts as an inhibitory factor to the vacancy agglomerate in a certain concentration.

## 2. Computational methods

We adopted a first-principles plane-wave method based on density functional theory (DFT) with the conjugated generalized gradient approximation and spin polarized Perdew-Burke-Ernzerhof pseudo-potentials to describe exchange-correlation interaction. According to the Vienna Abinitio Simulation Package code [23-25], the structure was simulated by using a 4×4×4 super-cell bcc W lattice, which contains 128 W atoms. 3.167 Å of the lattice parameter was given by the relaxed course and it is consistent with the experimental value. A 3×3×3 Monkhorst-Pack k-point mesh was used to perform the Brillouin Zone summation [26]. All calculations were performed with an plane wave energy cutoff of 400eV. All atoms are allowed to relax until all the atomic forces were smaller than $10^{-5}$eV/Å. The nudged elastic band (NEB) method was used to evaluate the maximum energy in the diffusion pathways [27].

To investigate the interaction between Re/Ta and vacancy, the binding energy need to be calculated, which is defined as:

$$E_{nVac,Re/Ta}^b = nE_{Vac} + E_{Re/Ta} - E_{nVac+Re/Ta} - nE_{perfect} \quad (1)$$

where $E_{Vac}$ and $E_{Re/Ta}$ are respectively the total energies of the system containing a vacancy and a substitutional Re/Ta, $E_{nVac+Re/Ta}$ is the total energy of the system containing $n$ vacancy and a substitutional Re/Ta atom, and $E_{perfect}$ is the total energy of a perfect W lattice consists of $N$ atoms without any vacancy.

For a pure W system with a mono-vacancy, the formation energy is simply defined as:

$$E_{Vac,per}^f = E_{Vac} - (\frac{N-1}{N})E_{perfect} \quad (2)$$

While for the system of containing transmutation elements Re/Ta system, the mono-vacancy formation energy is described a

$$E_{Vac,W-Re/Ta}^f = E_{Vac}^{W-Re/Ta} - E_{Re/Ta} + \frac{1}{N}E_{perfect} \quad (3)$$

where $E_{Vac}^{W-Re/Ta}$ denotes the total energy when the mono-vacancy and Re/Ta appear in the W system.

## 3. Results and discussion

### 3.1 Stability of mono-vacancy in W-Re /Ta system

The vacancy formation can be characterized by the binding energy and formation energy in the lattice. We consider four sites as the neighbors to the transmutation element, i.e., the first nearest neighbor (1NN), second nearest neighbor (2NN), third nearest neighbor (3NN) and fourth nearest neighbor (4NN). The structure is presented in the inset of Fig.1(a), where $W_1$, $W_2$, $W_3$ and $W_4$ denote the four nearest neighbors, respectively. The binding energy of Re/Ta and vacancy is regarded as a function of the distance between them. Fig.1(a) clearly shows that strong attractive interactions exist between Re and vacancies, consistent with the results in Ref.[28,29]; however, the interactions between Ta and vacancies are repulsive.

The mono-vacancy formation energies in pure W and W-Re/Ta system via different occupied positions are calculated and shown in Fig. 1(b). Firstly, contrast with the W-Re system, the $E_{Vac,W-Ta}^f$ for W-Ta system is significantly larger than 3.32eV of pure W [30]. Secondly, when facing of the same position, mono-vacancy formation energy of the W-Ta system is larger than that of the W-Re system, indicating that mono-vacancy is more likely to form in W-Re system than W-Ta system. The minimal $E_{Vac,W-Re}^f$ is 3.07eV occurred in 1NN site ($W_1$) in W-Re system and 3.39eV in 3NN site ($W_3$) in W-Ta system. Similar fluctuation in W-Ta system also appeared in others observation. The results suggest that the transmutation elements Re atom has a favored function to form the vacancy, while Ta acts as a delaying role to the vacancy via a solute drag effect [31,32].

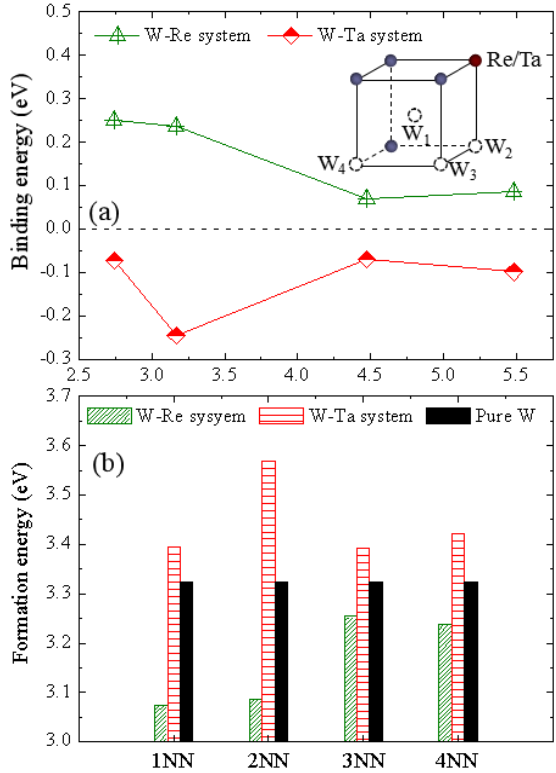

Fig. 1 (a) Binding energy between solute atoms and vacancy in W-Re/Ta system. The schematic is the W-Re/Ta lattice, the migration for the mono-vacancy is supposed from $W_1$ (1NN), $W_2$ (2NN), $W_3$ (3NN) and $W_4$ (4NN) sites, respectively. (b) Corresponding formation energy of mono-vacancy of pure W, W-Re and W-Ta system (eV), respectively.

### 3.2 Mono-vacancy diffusion in W-Re/Ta system

In order to reveal the law of the movements of point defects, together with different solute atoms, the mono-vacancy diffusion in W-Re/Ta system is assessed via the path models, as shown in Fig.2. There are four paths for a vacancy to diffuse in the neighborhood of a solute atom, labeled as $L_1$, $L_2$, $L_3$ and $L_4$, respectively. Meanwhile, opposite paths are labeled as $L_{2'}$, $L_{3'}$ and $L_{4'}$, respectively. The DFT-NEB method is adopted to calculate the migration barrier energy $E_m$.

Firstly, the vacancy diffusion energies along the paths $L_2$, $L_3$ and $L_4$ are respectively higher than the rightabout paths $L_{2'}$, $L_{3'}$ and $L_{4'}$, which can contribute to the stability of the whole system. In Fig. 2, the $E_m$ of pure tungsten is 1.67eV, agree with previous studies [33-35]. Obviously, the transmutation elements elements present some influence on the defect in W lattice. The minimum $E_m$ of the vacancy is 1.59 eV in $L_2$ of W-Re system and 1.43 eV in $L_1$ of W-Ta system, indicating that both Re and Ta can promote the mono-vacancy migration in W lattice. In contrast, Re atom has a smaller atomic radius than W atom, which will weaken the interatomic repulsive interaction. Secondly, the relative lower threshold energy in $L_2$ (1.59eV) makes the vacancy prefer to be replaced by W (not Re) in the W-Re system.

Therefore, Re atom can steady exist in W lattice and form solid solution with substrate W atom. Meantime, mechanical performance researches prove that Re can increase the malleability and hardness of substrate tungsten, and decrease the swelling under exposure [36, 37]. In contrast with Re, Ta is beneficial to move to a vacancy which ultimately promotes the vacancy's diffusion. Although the replacement atom is different, it is obvious that both Re and Ta atom manifest the positive effect to the mono-vacancy migration.

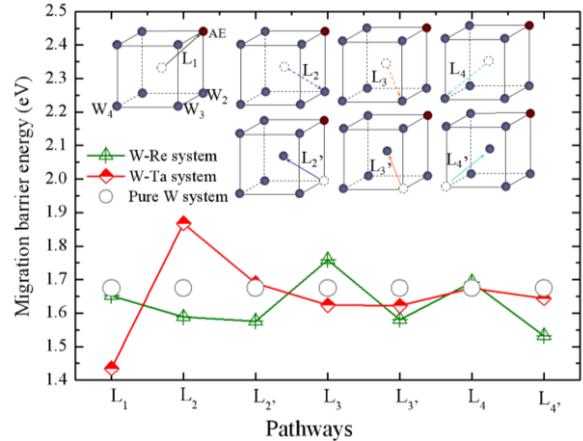

Fig. 2 One-step-jump migration energy for mono-vacancy, together with the diffusion of Re/Ta. There are four pathways from $L_1$ to $L_4$ and three rightabout ones as $L_{2'}$ to $L_{4'}$. Dashed circle denote vacancies, and TE refers Re or Ta atom, the transmutation element.

### 3.3 Clustering behavior of vacancy in W- Re/Ta system

An increasing thermal temperature would generate more vacancies in the lattice and lead to nucleation and growth of voids. Both vacancies and small clusters (including SIAs) can sink to the dislocations or grain boundaries. In a previous study, the presence of W seems to delay the SIAs via a solute drag force and therefore prevents the formation of vacancy in the W-doped Ta [21]. It reveals the influence of the solute atom can act as a positive or negative role to the point defect in the irradiation environment.

First, it is necessary to examine the clustering behavior of vacancy, together with the occurrence of Re/Ta nearby. The above calculations indicate that Re/Ta is benefit to the diffusion of mono-vacancy. In fact, the "Re-vacancy" binding energy shows an attraction between them. In contrast with Re, there is repulsion between "Ta-vacancy". These results have been further accessed via different size of vacancy clusters, where the transmutation elements Re and Ta were doped in the nearest neighborhood of the vacancy. The vacancy cluster [12] and complex configurations of "TE-V" structures are shown in Fig. 3.

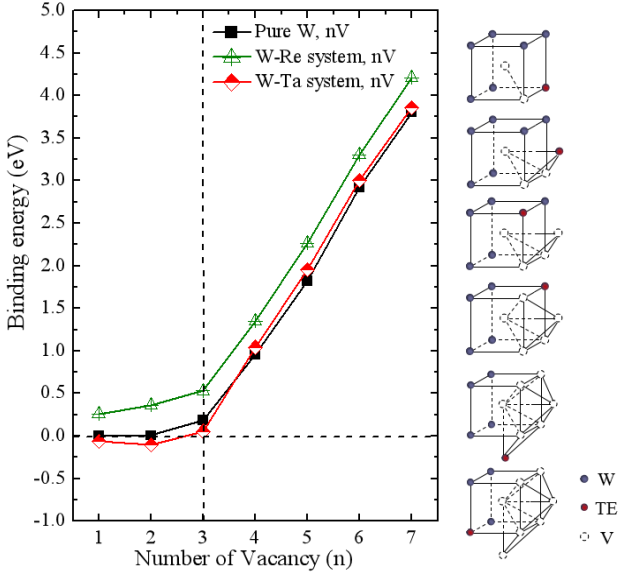

Fig. 3 Binding energy as a function of the number (*n*) of vacancy cluster in pure W, W-Re system and W-Ta system, respectively. The right are configurations of the "TE-V" system, with *n* from 1~7.

The cluster binding energies for six different vacancy configurations are calculated using *Eq.*(1) and shown in Fig. 3. Clearly, no matter in pure W or in W-Re/Ta system, the binding energy of vacancy increases with the increasing of vacancy number *n*, which indicates the vacancy cluster stability is closely related to its size. For pure W, the binding energy of divacancy has a small value of ~ 0.01eV. It shows there is a very weak attraction between the two vacancies, consistent with the results of the Bond-order potential calculations, and also supported by limited experimental value [38,39].

The TE induced influence is also presented in Fig. 3. Firstly, the presence of Re leads to a larger vacancy cluster binding energy than that of pure W. It indicates that Re contributes to the formation of the vacancy cluster and even for the nucleation and growth of void, which is consistent with the literature [28]. From this point of view, Re, acting as a kind of TE, is not favorite to the point defects in W system. Secondly, for the W-Ta system, it is easy to find that the binding energy presented transform behavior with the magnitude of the number *n* of the cluster. The substitution with Ta in W lattice brings both attractive and repulsion interaction in the vacancies. It becomes slightly repulsed when *n*<3, but strengthened the attraction when *n*≥3. This indicated that Ta would obstruct to form the diminutive vacancy cluster, and therefore the vacancy nucleation. On contrast, the solute Ta can promote to grow the larger cluster when the vacancy number *n* increased gradually.

Ref. [28] demonstrates that the vacancy formation depends on the volumetric strain, and the lattice strain plays a primary role to form vacancy cluster. Hence, the volumetric strain method was also adopted in this work to investigate the formation of vacancy cluster. The average reduced volume for a vacancy was defined as

$$\Omega_a = \frac{\Omega_{perfect} - \Omega_{cluster}}{n}, \quad (4)$$

where $\Omega_a$ is the average reduced volume of a vacancy, and $\Omega_{perfect}$ is the volume of the pure W. $\Omega_{cluster}$ is the volume of W system, containing the vacancy cluster, and *n* is the number of vacancy. The $\Omega_a$ and the vacancy binding energy are calculated and shown in Fig. 4.

In Fig. 4, the average reduced volume for the vacancy cluster $\Omega_a$ decreases with the increase of vacancy number *n*. The binding energy also presents an inverse relation with the relaxed volume $\Omega_a$. Fig. 4 reveals that an increasing attraction between vacancies is induced by the compression stress in lattice, which is caused by the decrease of the average reduced volume of vacancy.

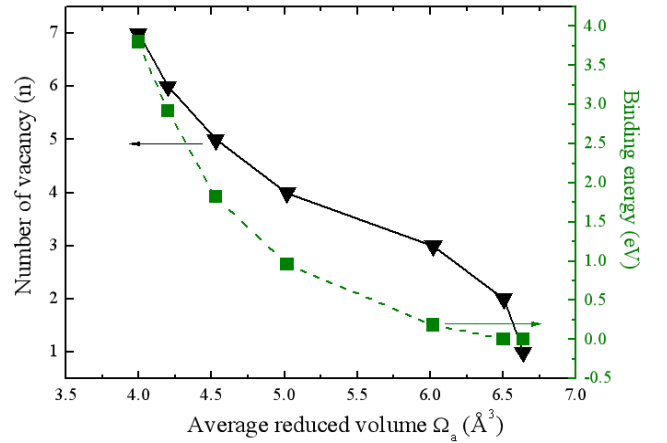

Fig. 4 Calculated vacancy binding energy and number *n* of vacancy as a function of the average reduced volume.

To better explore the vacancy clustering mechanism induced by transmutation elements, the whole system is supposed as two parts, i.e., "Vacancy-Vacancy" (the Pure W in Fig. 3), and "TE-Vacancy", respectively. The binding energy of "TE-Vacancy" is defined as

$$E^b_{W-Re/Ta} = E_{Re/Ta} + E_{nVac} - E_{Re/Ta+nVac} - E_{perfect} \quad (5)$$

The calculated $E^b_{W-Re/Ta}$ is shown in Fig. 5. The substitution with Re in W lattice can lead to the compressive strain at first, owing to the relatively smaller radius of Re atom. This compressive strain facilitates the stress between Re and vacancy, and therefore their binding energy is enhanced, as shown in Fig. 5. The binding energy increases gradually with the increase of the vacancy number. Accordingly, the doping of Re helps the nucleation and growth of vacancy (also agrees with Fig. 3). In contrast with Re, the lattice would expand with Ta's substitution since Ta has a bigger atom radius than W. Then a tensile strain brought by the local inflation can lead to a repulse between Ta atom and the vacancy. That's why the repulsive force occurred when there is a small cluster (n<3) in W-Ta system. However, $\Omega_a$ decreases with the continuously compressive lattice, together with the vacancy number *n* increased. At last, the tensile strain is balanced by the compressive process, and leads to an attraction between Ta and vacancy. This is also the reason for the turning point in Fig.5.

The results in Fig.5 can be used to interpret the TE influence on vacancy cluster behavior. Although the binding energy of "TE-Vacancy" is relatively weaker than that in "Vacancy-Vacancy" system (Fig. 3), we can conclude that Ta's existence delays the gathering of the point defects, when vacancy concentration is in a certain range. In these cases, the amount of Ta will also be an important factor to the vacancy clustering behaviors: the more the amount of Ta, the more the power of delaying vacancy clustering. In experimental, the damage level of the defect in Ta-10 at. % W is less than that in Ta-5 at. % W [22]. In other words, the calculations in our work can explain the experimental results in which the increased amount of Ta would promote the vacancy clustering as well as small agglomerate. TE's concentration would be a key factor in this process. These data can be used to deeper understanding evolution of the vacancy and vacancy cluster structure in the fusion environment.

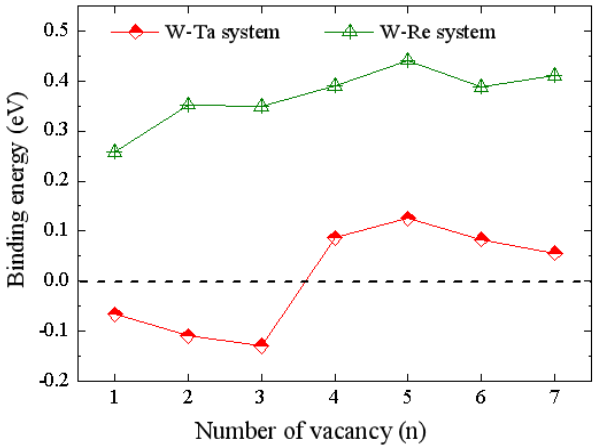

Fig.5 Binding energy of the "TE-Vacancy". Re can strengthen the interaction between Re and vacancies since the binding energy is always positive. On contract, Ta would repulse vacancies at small vacancy concentrations (n≤3) and attract vacancies at large vacancy concentrations (n≥4).

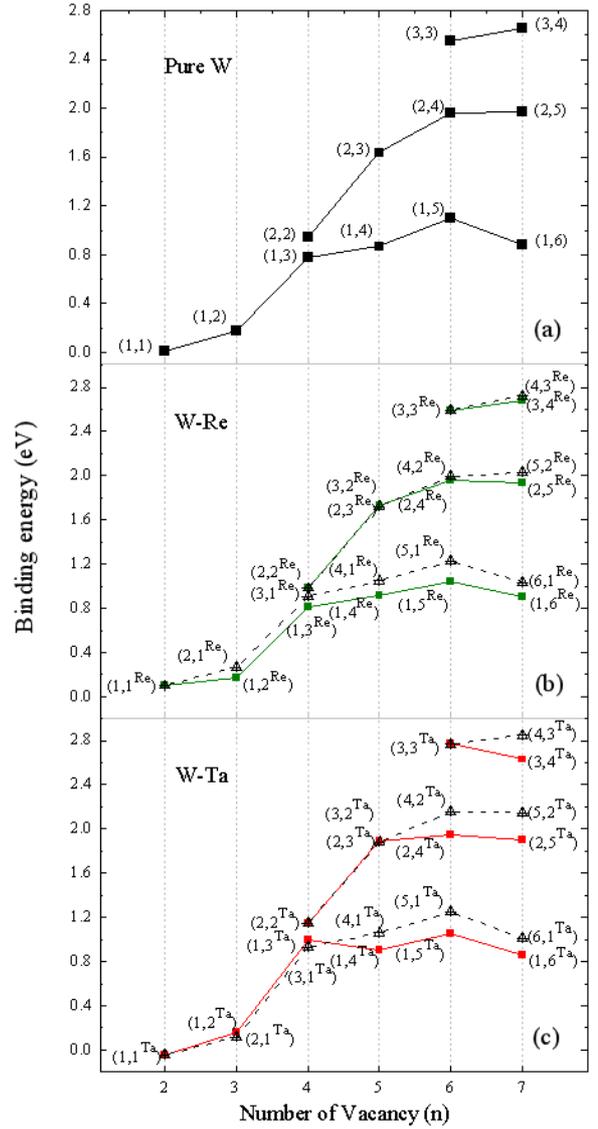

Fig.6 Binding energy in the vacancy dissociation (recombination) process in pure W and W-Ta/Re systems. Each vacancy cluster here is divided into two small parts according to the number (*n*) of vacancies in the cluster. Superscript (Ta or Re) represents that the TE atom locates in the corresponding part.

Kinetic Monte Carlo simulations suggest that Re-V can form small and stable Re-V clusters, which can further trap other vacancies to form larger clusters in low temperature [40], and can be fully dissolved at a high temperature of 2500 K [41]. In the following, we examined the interaction between vacancy cluster of different sizes in detail. For all examined systems, a larger vacancy cluster is divided into two small ones, such as (1, 1) when *n* = 2, (1, 3) and (2, 2) when *n* = 4, *etc*. The binding energy of pure W and Re-V clusters shown in Figs. 6(a) and 6(b) clearly manifests that the attraction between vacancy clusters gets stronger with the increase of the cluster size. Furthermore, the attraction between two clusters becomes larger when their size difference becomes smaller. For example, $E_{(3,4)} > E_{(2,5)} > E_{(1,6)}$, $E_{(4,3}^{Re}) > E_{(5,2}^{Re}) > E_{(6,1}^{Re})$, and $E_{(3,4}^{Re}) > E_{(2,5}^{Re}) > E_{(1,6}^{Re})$, where Re could favor to balance the volume of two different clusters, i.e., $E_{(4,3}{}^{Re}{}_{)} > E_{(3,4}{}^{Re}{}_{)}$, $E_{(5,2}{}^{Re}{}_{)} > E_{(2,5}{}^{Re}{}_{)}$, and $E_{(6,1}{}^{Re}{}_{)} > E_{(1,6}{}^{Re}{}_{)}$. Such energy differences also indicate that a vacancy cluster is favorable to grow by combining another cluster relative to a mono-vacancy. On the other side, releasing a single vacancy is likely to be the main mechanism of the vacancy cluster dissociation.

Contrast to "Re-Vacancy" cluster, the mechanism of vacancy growth of "Ta-Vacancy" cluster changes due to the repulsion between Ta and vacancy when $n$ is small ($n \leq 4$). For example, $E_{(2,1}{}^{Ta}{}_{)}$ and $E_{(3,1}{}^{Ta}{}_{)}$ become slightly lower than $E_{(1,2}{}^{Ta}{}_{)}$ and $E_{(1,3}{}^{Ta}{}_{)}$, respectively [see Fig. 6(b)]. This feature is consistent with the performance of W-Ta system in Fig 3.

## 4. Conclusion

Transmutation elements Re and Ta effects on vacancy characterization have been assessed in tungsten by first-principles method. The formation and diffusion of vacancy as well as small clusters behaviors were examined. The influences induced by transmutation atom Re and Ta were investigated in detail. The binding energy showed a strong attraction existed between Re and vacancy, while Ta presented a repulsive one. Mono-vacancy is more likely to form in W-Re system than pure W and W-Ta systems. However, both Re and Ta atom promoted the mono-vacancy migration. Nevertheless, Re contributed to the nucleation and growth of the vacancy cluster so as to be unsuitable for a large amount in W material. The substitution of Ta in W lattice would obstruct to the nucleation of diminutive vacancy cluster when it is within a certain range of Ta concentration. A vacancy cluster is more favorable to grow by combining another cluster relative to a mono-vacancy, while its dissociation is more likely to be a process of releasing a single vacancy. Such aggregation/dissociation behaviors can be understood from the close relation between the average reduced volume and the compression stress in W-Re/Ta system. Different from Re, Ta is expected to enhance the combination of vacancy and SIA according to its obstruction to the vacancy nucleation.

**Acknowledgement：**

The research is supported by the National Key R&D Program of China (Grant No.2018YFE0308102, 2017YFE0301404), National Natural Science Foundation of China (Grant No.11405138,11647108).